# Efficient LLM inference solution on Intel GPU

Hui Wu, Yi Gan, Feng Yuan, Jing Ma, Wei Zhu, Yutao Xu,
Hong Zhu, Yuhua Zhu, Xiaoli Liu, Jinghui Gu, Peng Zhao

Intel® Corporation

{hui.h.wu, yi.gan, feng1.yuan, jing1.ma, wei2.zhu, hong.zhu, yuhua.zhu, xiaoli.liu, jinghui.gu, patric.zhao}@intel.com

**Abstract**

Transformer based Large Language Models (LLMs) have been widely used in many fields, and the efficiency of LLM inference becomes hot topic in real applications. However, LLMs are usually complicatedly designed in model structure with massive operations and perform inference in the auto-regressive mode, making it a challenging task to design a system with high efficiency.

In this paper, we propose an efficient LLM inference solution with low latency and high throughput. Firstly, we simplify the LLM decoder layer by fusing data movement and element-wise operations to reduce the memory access frequency and lower system latency. We also propose a segment KV cache policy to keep key/value of the request and response tokens in separate physical memory for effective device memory management, helping enlarge the runtime batch size and improve system throughput. A customized Scaled-Dot-Product-Attention kernel is designed to match our fusion policy based on the segment KV cache solution. We implement our LLM inference solution on Intel® GPU and publish it publicly. Compared with the standard HuggingFace implementation, the proposed solution achieves up to 7x lower token latency and 27x higher throughput for some popular LLMs on Intel® GPU.

## 1. Introduction

Recently, Transformers [1] based Large Language Models (LLMs) [2-4] have received widespread attention. With the capability of content understanding and generation, LLMs have been applied in many downstream applications [5-9]. However, LLMs are usually complicated designed, growing larger [10] and deeper [11], making it a challenging task to achieve a highly efficient LLM inference system.

LLM inference systems are usually used in different scenarios like the latency-critical online and throughput-oriented offline applications [12, 13]. For online serving, the latency indicates the time consumption of generating some tokens, reflecting the online system fluency, the lower latency the better user experience. For offline application, the throughput indicates the number of tokens generated at a time, reflecting the system resource utilization ratio, the higher value of the throughput the lower cost of the system.

For online mode, it is critical to lower the latency and improve the response efficiency for request. However, the deep decoder layers combined with multiple operations make it not easy to achieve low latency for a single token. Taking Llama2 [14] for example, the basic structure is shown in Figure 1, containing multiple decoder layers with several modules in each, like Linear, Rotary Position Embedding (RoPE), Scaled Dot Product Attention (SDPA), Root Mean Square Layer Normalization (RMSNorm) and Activation. Simplifying the model structure and reducing the number of operations in LLM structure can help lower the latency.

Throughput criteria should also be considered to reduce cost. Enlarging the batch size value and improving the occupancy of computation resource can help achieve high throughput. However, the auto-regressive principle makes LLM inference a memory consuming system, limiting the batch size value and the throughput that can be achieved on HPC platform with constraint device memory. Generally, during LLM inference process, the input tokens (prompt) will be computed at first in prefill phase and the output tokens (response) will be generated step by step in decoding phase based on the previous generated tokens until the terminated token being achieved. At each time step in the decoding phase, some candidate tokens will be generated at a time, and beam search method [15] is usually applied to select some tokens with high confidence scores. With this auto-regressive principle, the key/value of each decoder layer at each time step are kept on device memory during the whole decoding phase (named as KV cache policy [13]), consuming more and more memory with time step increasing. It is critical to effectively manage the device memory to enlarge the batch size and improve the throughput.

In this paper, we design and implement a LLM inference solution with low latency and high throughput on Intel® GPU, currently covering some of the most popular LLMs including GPT-J [16], Llama [4] / Llama2 [14], OPT [18], and Bloom [19]. The main contributions of this paper include:
1. We propose an efficient LLM inference solution and implement it on Intel® GPU. The implementation is available on-line with our [Intel®-Extension-for-Pytorch repository](#).
2. To lower latency, we simplify LLM decoder layer structure to reduce the data movement overhead. In addition, we design a deep fusion policy to fuse both GeMM and Element-wise operations as more as possible. For some popular LLMs mentioned above with parameter sizes from 6B ~ 176B, our inference solution achieves up to 7x lower token latency compared with the standard HuggingFace implementation ([v4.31.0](#)) [20].
3. To improve throughput, we propose a segment KV cache policy to keep prompt and response key/value into discrete device memory, making the prompt key/value being shared by different response tokens to avoid memory wasting. On the same device configuration, the throughput (tokens per second) of our inference solution is up to 27x higher than that of the standard HuggingFace implementation for the popular LLMs with parameter sizes from 6B ~ 176B.
4. Based on our fusion policy and segment KV cache methodology, we design a highly efficient kernel to fuse all computation steps in SDPA module together with possible index selection process for beam search usage.

## 2. Related Works

The efficiency of Transformers based workloads are usually bottlenecked by memory access [21, 22], where reading and writing data account for a large portion of runtime. Reducing memory-bound operations overhead in LLMs can help improve efficiency. Kernel fusion is a general approach to reduce such overhead by combining several successive operations together and compute them in a single kernel to avoid frequent memory access to slow GPU high bandwidth memory (HBM). Some previous works [23-26] focused on element-wise operations fusion to reduce kernel launch and memory access overhead. Beyond general fusion on element-wise operations, Fang et al. [26] proposed TurboTransformers to fuse Element-wise and Reduction operations between two GeMMs together. Aminabadi et al. [12] designed some customized GeMM kernels based on the characteristics of Transformer related workloads and adopted a fusion policy to combine operations of data layout conversion and reduction together with the customized GeMMs. Dao et al. [21] focused on improving the SDPA module efficiency and proposed a FlashAttention algorithm to combine all computation steps (Batch-GeMM, Softmax, possible Masking, Batch-GeMM) together, achieving up to 3x faster than the standard attention implementation in different user scenarios. Based on FlashAttention [21], FasterTransformer [27] further fused the index selection for beam search into the SDPA kernel to further remove the data movement operations. Considering LLMs are weight bounded workloads, compression techniques including pruning [28], quantization [29, 30], knowledge distillation [17] and low-rank factorization [31] are usually applied to compress the LLM weight so that to reduce the memory access sizes and improve efficiency. [29, 30] focused on LLM weight quantization and applied low precision Matmul Multiplication for feed-forward and attention projection layers in transformers to reduce the inference memory usage. Consider irrelevant information will be extracted with long context, [32] adopt sparse transformer to explicitly extract the most relevant segments in attention module to reduce the computation.

Many related works also focus on improving the throughput of LLM inference system. Batching is an important technique to enhance hardware resource utilization and improve throughput. However, LLM inference is memory consuming, limiting the batch size value can be set on hardware platform with constraint device memory. Effective memory management is critical to help enlarge batch size and improve throughput. Due to the auto-regressive principle, KV cache in LLM consumes large device memory, larger than 30% device memory for a 13B-parameter LLM [33]. To effectively manage the KV cache, Kwon et al. [33] proposed the PagedAttention technique to map logical contiguous key/value to separate psychical memory, achieving near-zero memory wasting and making flexible KV cache sharing across different requests. Generally, prompt requests from users are in different sequence length, static batching (padding different length requests to the max length in a batch) will waste hardware resource since requests finished earlier cannot be sent back to the client until the whole batch finished. To fix this issue, ORCA system [34] proposed an iteration-level scheduling mechanism to invoke the execution engine running only a single timestep of the model on the batch so that the finished tokens can be sent back immediately.

Moreover, some LLM inference engines like TVM [35] and ONNXRumtime [24] are designed to provide highly efficient serving on different hardware platforms and improve the system efficiency with optimized kernels based on different hardware characteristics. Besides inference on single GPU card, DeepSpeed [12] also considers extending the LLM inference on multiple GPU cards and proposes tensor/pipeline/expert parallelism techniques.

## 3. Proposed Method

In this paper, we design an efficient LLM inference solution and implement it on Intel® GPU. To lower the latency, we simplify the structure of Transformer decoder layer by reducing data movement operations and applying customized kernel fusion policy. To more effectively manage the device memory and improve throughput, we propose a segment KV cache algorithm to make prompt key/value being shared between different response tokens. A customized SDPA kernel is designed to support our fusion policy and segment KV cache algorithm.

### 3.1 Model Structure Simplification

As we mentioned above, LLMs are usually complicatedly designed with multiple decoder layers consisting of massive operations to capture context information. The decoder layer usually has two basic modules Multi Head Attention (MHA) and Feed-Forward (FF), which are connected by some normalization operations. Taking Llama2 for example, the basic model structure is shown in Figure 1. At time step $t$, in MHA module, three Linear operations are firstly adopted to generate **$Query_t$**, **$Key_t$**,

**Value**$_t$, followed by possible RoPE for position embedding. Then SDPA module is applied for attention context computation, and finally another output Linear is utilized for feature projection. Before the SDPA module, some data movement operations including Transpose, Cat and Index Select are used. The Transpose operation is applied on **Query**$_t$, **Key**$_t$, **Value**$_t$ for data layout transformation from [BS × BW, N$_t$, H, D] to [BS × BW, H, N$_t$, D], where BS, BW indicate the runtime batch size, beam width value, and H, D indicate the number of attention head and head dimension of the model, while N$_t$ indicates the sequence length at time step $t$. Then past key/value (consisting of prompt **Key**$_0$/**Value**$_0$ and response **Key**$_{1\sim t-1}$ / **Value**$_{1\sim t-1}$) are extracted by Index Select operations based on the **Indice**$_t$ tensor generated from Beam Search module, which will be concatenated with the current **Key**$_t$/**Value**$_t$ to create KV cache for attention context computation in SDPA module. The generated attention context in memory layout of [BS × BW, H, N$_t$, D] will be converted back to [BS × BW, N$_t$, H, D] for final Linear projection.

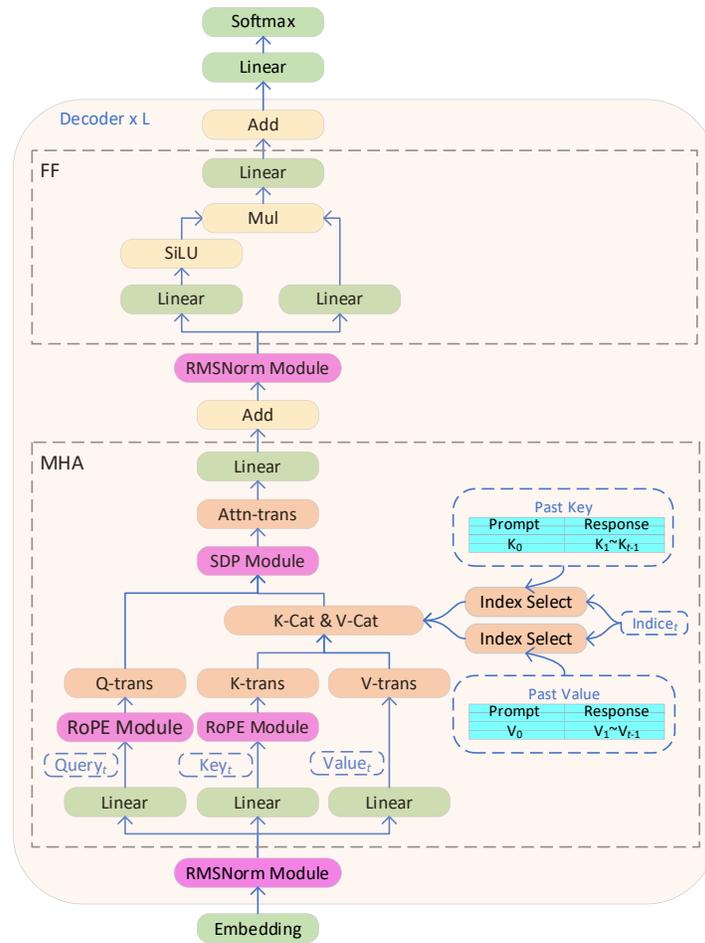

Prefill phase & Decoding phase

Figure 1. The standard flowchart of LLM inference with Llama2.

The data movement operations in MHA like Transpose, Cat, Index Select labeled in orange in Figure 1 will cause memory access problems, such as poor cache locality, and introduce kernel launch overheads. In this paper, we simplify the MHA structure by reducing these operators as shown in Figure 2. At prefill phase, **Query**$_0$, **Key**$_0$, **Value**$_0$ in memory layout of [BS × BW, N$_0$, H, D] (named as batch first layout) are generated, directly followed by a customized designed SDPA kernel for attention context computation without any transformation operation. The customized designed SDPA kernel accepts input and generate output tensors in batch first layout ($[BS \times BW, N_0, H, D]$) to avoid the transformation overhead, as shown in Figure 2(a). At decoding phase, Cat operations shown in Figure 1 are used to combine key/value generated at different time step together. To remove the Cat overhead, we pre-allocate KV cache buffers for response tokens with sequence length $N_{step}$ in memory layout of $[N_{step}, BS \times BW, H, D]$ (named as sequence first layout). At time step $t$ of the decoding phase, the $\mathbf{Key}_t$ and $\mathbf{Value}_t$ are kept in corresponding parts of the pre-allocated buffers $[N_{t-1}: N_t, BS \times BW, H, D]$. The sequence first layout ensures the $\mathbf{Key}_t$ and $\mathbf{Value}_t$ are always contiguously arranged for easy computation unnecessary to aware of the hyper-parameter $N_{step}$. Then the prompt $\mathbf{Key}_0$, $\mathbf{Value}_0$ and the response key/value ($\mathbf{Key}_{1\sim t}$, $\mathbf{Value}_{1\sim t}$ in the pre-allocated KV cache buffers) together with the $\mathbf{Indice}_{1\sim t}$ tensor (generated by Beam Search module) will be used by our customized SDPA kernel to compute attention context in memory layout of $[1, BS \times BW, H, D]$ for final linear computation without any transformation. To align the sequence first layout policy for KV cache without introducing extra data movement overhead, we convert the input tensor (named as Hidden States in Figure 2) before the first decoder layer from batch first layout to sequence first layout. The sequence first layout will be propagated internal and external all the decoder layers until arriving the last decoder layer, the output of which will be converted back to batch first layout. The optimized MHA structure can be seen in Figure 2, illustrating that all data movement operations labeled in orange in Figure 1 are removed.

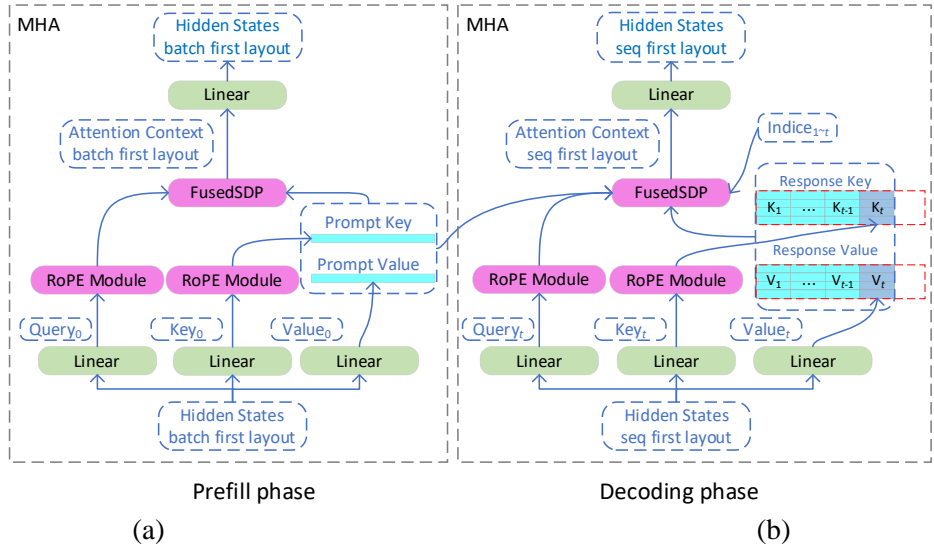

Figure 2. Optimized MHA module in Llama2 (a) at prefill phase, and (b) decoding phase, respectively.

Moreover, we respectively fuse the multiple operations in RMSNorm, RoPE and SDPA modules (labeled in pink in Figure 1) to single kernels. The three Linear operations (respectively generating query/key/value) are also combined into a single one and the element-wise operations labeled in yellow in Figure 1 are further fused with their previous Linear operations. The optimized Llama2 structure is shown in Figure 3. Compared with the original structure shown in Figure 1, the data movement operations labeled in orange and the element-wise operations labeled in yellow are all removed. The modules labeled in pink containing multiple operations are fused into single kernels. Even we introduce extra sequence/batch first layout transformation operations at decoding phase, they happen only once before the first and last decoder layers at each time step, almost no overhead compared with the whole inference. With our optimization, the number of operations in each Llama2 decoder layer has been reduced to nine, much smaller than the original hundreds of operations.

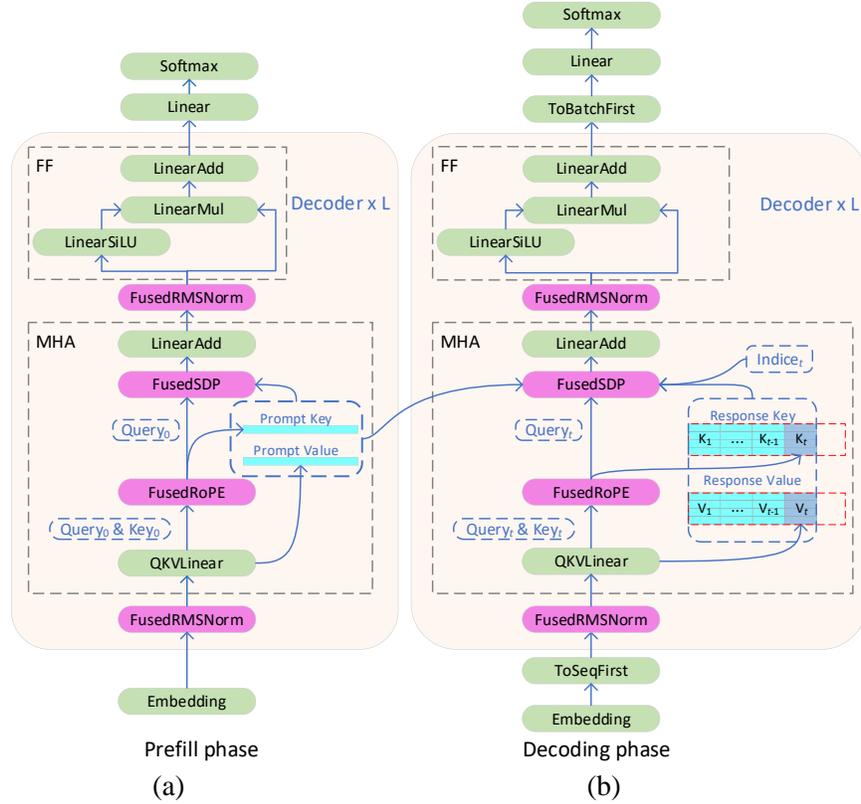

Figure 3. Optimized flowchart of LLM inference with Llama2 (a) at prefill phase, and (b) decoding phase, respectively.

*3.2 Segment KV Cache*

LLM inference is memory consuming caused by large parameter size and KV cache policy, limiting the batch size value and then further impacting the system throughput. Assuming the LLM has L numbers of decoder layers, H attention heads, D head dimension, then the KV cache size of one token can be computed as Eq (1), where 2 indicates the key and value. The runtime batch size, beam width are BS, BW, and the total sequence length is combined with prompt sequence length $N_{prompt}$ and response sequence length $N_{response}$. In standard KV cache implementation, the prompt and response key/value will be concatenated together to create contiguous KV cache in shape of $[BS \times BW, N_{prompt} + N_{response}, H, D]$. Then the KV cache size at the last time step can be computed as Eq (2). In standard implementation, keeping prompt and response key/value in contiguous KV cache buffers will waste device memory since the prompt key/value should be extended BW times. Moreover, since the KV Cache buffers grow larger at each time step in decoding phase, new bigger buffers will be allocated while may not reuse the previous smaller KV Cache buffers, resulting in many memory fragments. Considering the extreme situation, all the physical device memory of KV cache at each time step cannot be reused, then the total memory fragment should be the sum of KV Cache at each time step, much larger than the KV cache itself at the last time step.

$$\text{cache}_{\text{token}} = 2 \times L \times H \times D \times \text{sizeof(datatype)} \tag{1}$$

$$\text{cache}_{\text{standard}} = BS \times BW \times (N_{\text{prompt}} + N_{\text{response}}) \times \text{cache}_{\text{token}} \tag{2}$$

To avoid keeping duplicated prompt key/value and try best to reduce the memory fragment, we propose a segment KV cache policy, holding the prompt and response key/value in different buffers and regularly empty cache the fragments at each time step. At prefill phase, the prompt key/value in shape of $[BS, N_{\text{prompt}}, H, D]$ for each decoder layer are generated and kept on the device as shown in Figure 3(a). Then, at decoding phase, we pre-allocate the response KV cache in shape of $[N_{\text{step}}, BS \times BW, H, D]$ for each decoder layer. The response key/value at each time will be kept on the corresponding part of KV cache, as shown in Figure 3(b). With the proposed segment KV cache policy, the prompt key/value can be share with different response tokens.

In some previous works, $N_{\text{step}}$ is usually set as the max position length of the model like 4096 in Llama2, which can help reduce the memory fragments while resulting in memory wasting if the runtime response has smaller sentence length. To fix this issue, we set $N_{\text{step}}$ value dynamically increasing with step = 16. Taking response length larger than 16 for example, at the 1$^{\text{st}}$ timestep ($N_{\text{response}} = 1$), we pre-allocate the KV cache with $N_{\text{step}} = 16$. Then at the 17$^{\text{th}}$ timestep ($N_{\text{response}} = 17$), $N_{\text{step}}$ will be set as 32 and new KV cache will be allocated, the corresponding part of which will be fulfilled with the original KV cache data and the other part will keep the new key/value data. At the same time, the previous KV cache with $N_{\text{step}} = 16$ will be manually empty cached to improve the probability of memory reuse. In this way, the length of the response KV cache will increase with step = 16, reducing some data movement operations of each time step while not introducing much memory wasting or fragments. The memory consumption of the segment KV cache policy with $N_{\text{step}}$ increasing with step = 16 can be computed as Eq (3).

$$\text{cache}_{\text{segment}} = BS \times (N_{\text{prompt}} + BW \times \text{Ceil}(\frac{N_{\text{response}}}{\text{step}}) \times \text{step}) \times \text{cache}_{\text{token}} \tag{3}$$

Table 1. Different LLM Configurations

| Model | Decoder Layer (L) | Attention Heads (H) | Head Dim (D) |
|---|---|---|---|
| GPT-J-6B | 32 | 32 | 128 |
| Llama2-13B | 40 | 40 | 128 |
| OPT-30B | 48 | 56 | 128 |
| Bloom-176B | 70 | 112 | 128 |

Taking LLMs with different parameter sizes from 6B to 176B as shown in Table 1 for example, the runtime BW, $N_{\text{prompt}}$ and $N_{\text{response}}$ set as 4, 1024, 1024, respectively. Not considering the memory fragment, the KV cache memory consumption in Float16 datatype with different BS values at the last time step are shown in Figure 4, where standard and segment represent the standard KV Cache policy and our proposed segment KV cache policy, respectively. Taking GPT-J-6B for example, with BS = 32, the proposed segment KV cache consumes 86GB device memory, only about 63% of the standard KV cache policy with 137GB, saving 51GB device memory in total. Thus, for GPU hardware with constraint device memory, the proposed segment KV cache policy consumes less memory, so that larger batch size value can be adopted to improve system throughput.

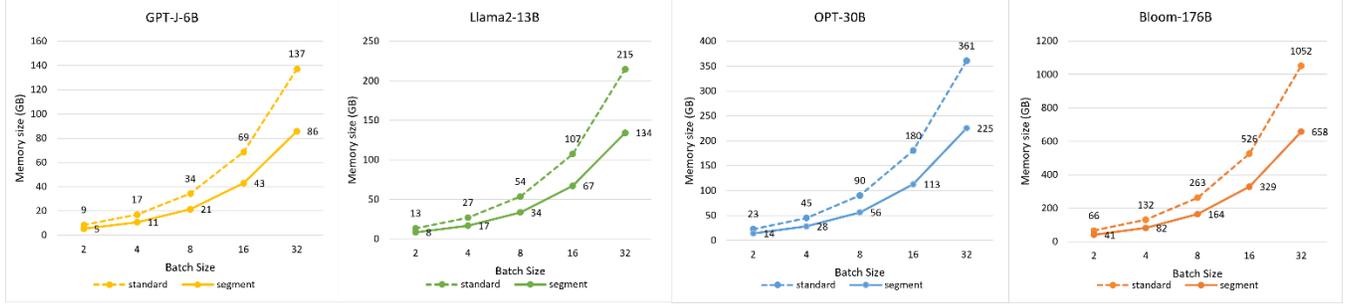

Figure 4. Memory consumption of standard KV cache and segment KV cache on different LLMs

## 3.3 Customized SDPA Kernel

Inspired by FlashAttention [21] and FasterTransformer [27], we fuse all computation steps in SDPA module as well as the possible Index Select operations into a single kernel based on our segment KV cache policy. For the customized SDPA kernel at decoding phase, the input tensors of query, prompt key/value and response key/value are in shapes of $[1, BS \times BW, H, D]$, $[BS, N_{prompt}, H, D]$ and $[N_{response}, BS \times BW, H, D]$, respectively. The output tensor of attention context is in shape of $[1, BS \times BW, H, D]$. In our implementation, BS and H are paralleled in different GPU work groups (blocks). In a single GPU work group, the input tensors of query $\mathbf{Q}$, prompt key/value ($\mathbf{K_{prompt}}$, $\mathbf{V_{prompt}}$) and response key/value ($\mathbf{K_{response}}$, $\mathbf{V_{response}}$) are in shapes of $[1, BW, D]$, $[N_{prompt}, D]$ and $[N_{response}, BW, D]$, respectively. The output tensor of attention context $\mathbf{O}$ is in shape of $[1, BW, D]$. The customized SDPA kernel at prefill phase is a special case of that at decoding phase, only using input tensors of query and prompt key/value for attention context computation. We take the computation in a single work group of SDPA kernel at decoding phase as an example to elaborate how we implement the kernel.

*Index select fusion*. The beam indices tensor $\mathbf{Indice_{1\sim t}}$ in shape of $[BW, N_{response}]$ is obtained in the same way as that in FasterTransformer [27]. Based on $\mathbf{Indice_{1\sim t}}$, the actual response key/value on each beam position at different time step are selected. Taking the 2[nd] beam position at the 1[st] time step for example shown in Figure 5(a), the indices value at this position is 3, then the response key value at this position should be the 3[rd] item at 1[st] time step in key cache.

*Kernel implementation*. In a single work group, $\mathbf{Q}$ will be loaded into register at first. Then we for loop to load the corresponding parts of $\mathbf{K_{prompt}}$ and $\mathbf{V_{prompt}}$ at $N_{prompt}$ dimension to compute the attention context like FlashAttention. In next step, we apply the index select process as mentioned above on KV cache and for loop to load the corresponding part of $\mathbf{K_{response}}$ and $\mathbf{V_{response}}$ at both BW and $N_{response}$ dimensions for attention context computation. Finally, the attention context value respectively computed based on prompt and response key/value will be accumulated together to get the result and be written into the HBM. The corresponding flow can be seen in Figure 5(b).

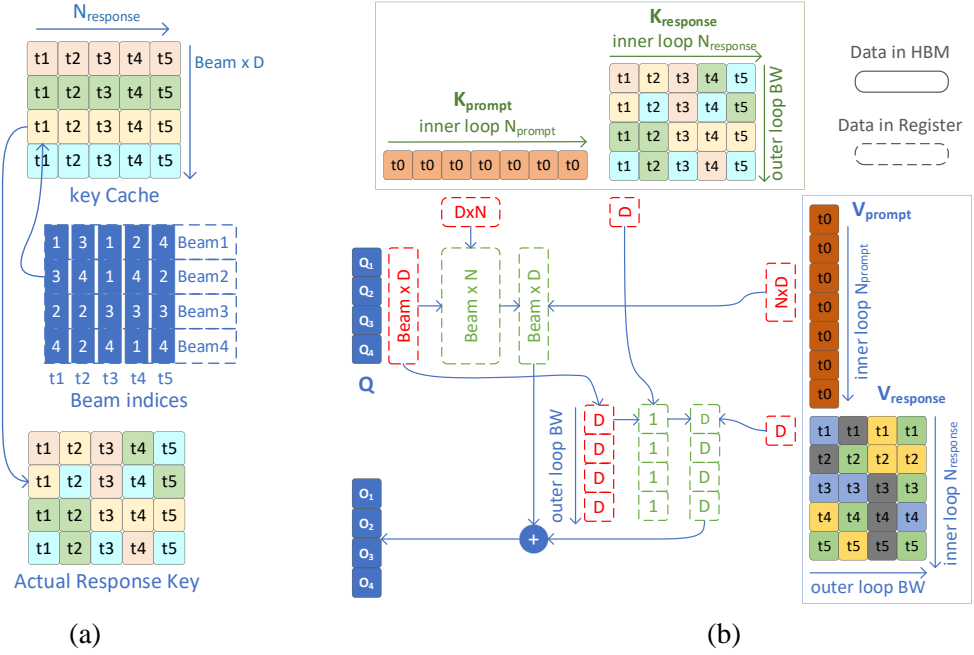

Figure 5. Implementation of customized SDPA kernel. (a) Index select; (b) Customized SDPA kernel based segment KV cache policy.

## 4. Experiments

We implement our LLM inference solution on Intel® GPU and perform the experiments on a cluster of 4 × Intel® Data Center Max 1550 GPU cards with 2 Tiles per Card, 64 $X^e$-cores & 512 EUs per Tile. The device memory per Tile is 64GB with effective memory bandwidth about 1000GB/s. These GPUs are hosted on a 2x Intel® Xeon® 8480+ system running Ubuntu 22.04.3. Our optimization solution code is released in [Intel®-Extension-for-Pytorch (v2.1.10)](#) and our software stacks are publicly available [here](#) for performance reproduction.

In our experiments, if no specification, we will use the configuration of Float16 datatype, input prompt length $N_{prompt} = 1024$ and output response length $N_{response} = 128$ for performance evaluation. At first, we perform experiments on our customized SDPA kernel to show the kernel efficiency. Then we apply our inference solution on some popular LLMs listed in Table 1. For models with small parameter sizes GPT-J-6B and Llama2-13B, we run them on Max 1550 GPU single tile. For models with large parameter sizes OPT-30B and Bloom-176B, we respectively run them on Max 1550 GPU one card two tiles and four cards eight tiles. For the large parameter size models, we use automatic Tensor Parallel (autoTP) [12] algorithm to split model layers horizontally to run them across GPU devices.

We use both latency and throughput criteria for performance evaluation. The latency includes two ratios: first token latency indicating the latency of the first token generation in a batch, and next token latency indicating the average latency of the second to last token generation in a batch. For inference throughput evaluation, we figure out the largest batch size can be set and collect the corresponding latency. Then the throughput can be computed as Eq (4).

$$\text{Throughput} = (BS_{max} \times N_{response})/\text{latency} \tag{4}$$

*4.1 SDPA Kernel Performance Evaluation*

We implement the customized SDPA kernel with the input tensors of prompt key/value and response key/value in shapes of $[BS, N_{prompt}, H, D]$ and $[N_{response}, BS \times BW, H, D]$, respectively. Taking Llama2-13B ($H = 40$ and $D = 128$) with BW=4 for example, the effective memory bandwidth values of the customized SDPA kernel with different BS are shown in Figure 6. For $BS = 1$, our customized kernel (prompt with batch first layout and response with sequence first layout) achieves 727GB/s, only about 72% of the peak memory bandwidth. There are 64 $X^e$-cores on Max 1550 GPU single tile, and at least 64 numbers of work groups are needed to fulfill all the $X^e$-cores, otherwise only low hardware resource occupancy and poor efficiency can be achieved. In our implementation, BS and H are parallelly computed in different work groups. With $BS = 1$, the paralleled workgroup number is $H = 40$, not fulfilling the total 64 $X^e$-cores. Small workload size with $BS = 1$ will waste hardware resource and results in poor efficiency. When BS becomes larger and equals 16, $BS \times H = 640$ workgroups are scheduled on 64 $X^e$-cores and we achieve high hardware resource occupancy with memory bandwidth 942GB/s. The kernel efficiency will decrease with BS continuing becoming larger, because for response key/value we for loop $N_{response}$ to load D elements with stride $BS \times BW$, the bigger the stride value, the lower probability of cache hitting.

We also perform experiment to compare SDPA kernel efficiency with prompt key/value in batch first (proposed method) and sequence first layout (aligned with response KV cache layout). As shown in Figure 6, for $BS = 1$, the two layouts can both be simplified as $[N_{prompt}, H, D]$ and achieve similar memory bandwidth value about 720GB/s. For larger BS, the prompt key/value in batch first layout $[BS, N_{prompt}, H, D]$ with stride H has higher cache hit probability than that in sequence first layout $[N_{prompt}, BS, H, D]$ with stride $BS \times H$, so that our customized SDPA kernel implemented with prompt key/value in batch first layout (different with KV cache layout) can achieve better performance efficiency.

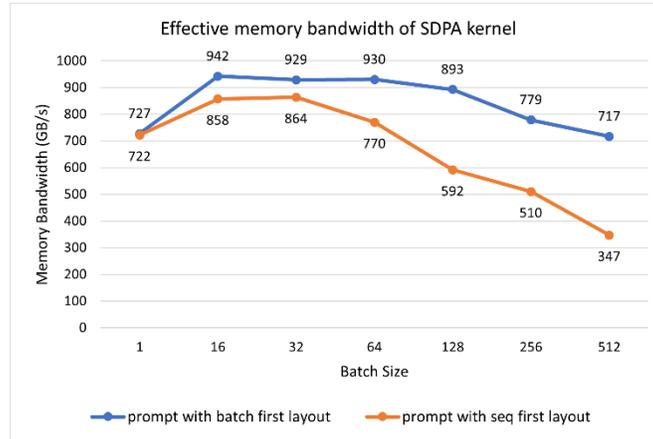

Figure 6. SDPA kernel performance with different batch sizes for different prompt key/value memory layout

### 4.2 System Latency Evaluation

4.2.1 Latency evaluation with different prompt lengths

Firstly, we perform experiments on Llama2-13B to figure out the influence of the prompt sequence length $N_{prompt}$ on the latency with response tokens $N_{response} = 32, BS = 1, BW = 4$. We collect the first token latency and the next token latency with different $N_{prompt}$ from 32 to 4096. Performance results of the standard HuggingFace (named as HF) and our proposed solution (Proposed) are shown in Figure 7, among which the HF implementation will out of memory for $N_{prompt} = 4096$. From Figure 7(a) we can see that, with $N_{prompt}$ becomes longer, the first token latency of HF implementation becomes higher. $N_{prompt} = 128$ is the inflection point, smaller than which the latency increases slowly with low hardware occupancy, while 2 times increasing for $N_{prompt}$ larger than 128, almost achieving full hardware resource occupancy. Compared with HF implementation, the proposed method accesses only 1/BW amount data in HBM, and the corresponding inflection point happens at $N_{prompt} = 512$, BW times longer than 128. Moreover, the proposed method achieves much lower first token latency than standard HF implementation for any $N_{prompt}$ values, and more than BW times when $N_{prompt}$ larger than 512 with full hardware occupancy. Figure 7(b) shows the same situations that the proposed method achieves much lower next token latency with any different prompt sequence length.

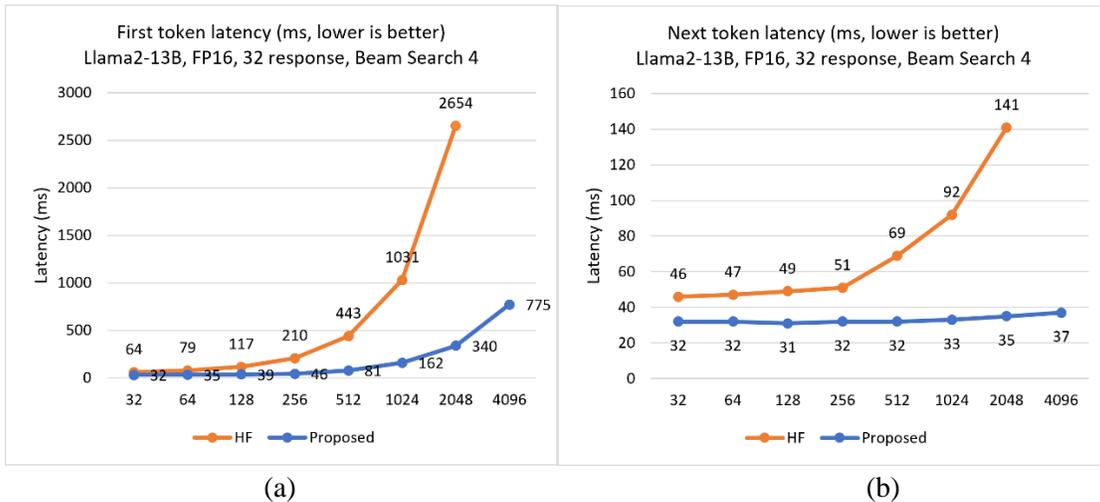

Figure 7. Llama2-13B latency of (a) the first token, and (b) the next token with different prompt sequence length, respectively.

4.2.2 Latency evaluation on different LLMs

We also perform experiment to evaluate the performance influence of different BW values (BW = 1 and 4) with BS = 1. The performance results of the first and next token latency on different LLMs are shown in Figure 8. Both the first and next token latency of the standard HF implementation grows linearly with BW becomes larger. However, the BW value has little impact on the proposed method, with almost the same first and next token latency with BW 4 and 1 as shown in Figure 8.

Moreover, we compare the token latency of the proposed method and standard HF implementation. For BW 1, the first and next token latency of the proposed method achieves about 1.1x ~2x faster than the standard HF implementation on different LLMs. The benefits come from our deep fusion policy with highly efficient kernels, removing data movement operations and fusing element-wise operations, which can help reduce both the HBM access and kernel launch overhead. For BW 4, the first token latency values of our proposed method are 4x ~7x lower than the standard HF implementation. Besides the benefits from deep fusion policy, the segment KV cache method (unnecessary to extend the prompt with BW = 4 times) helps reduce both memory access and kernel computation costs. The performance of next token latency of the proposed method is also much lower than the HF implementation especially for BW = 4 as shown in Figure 8(b).

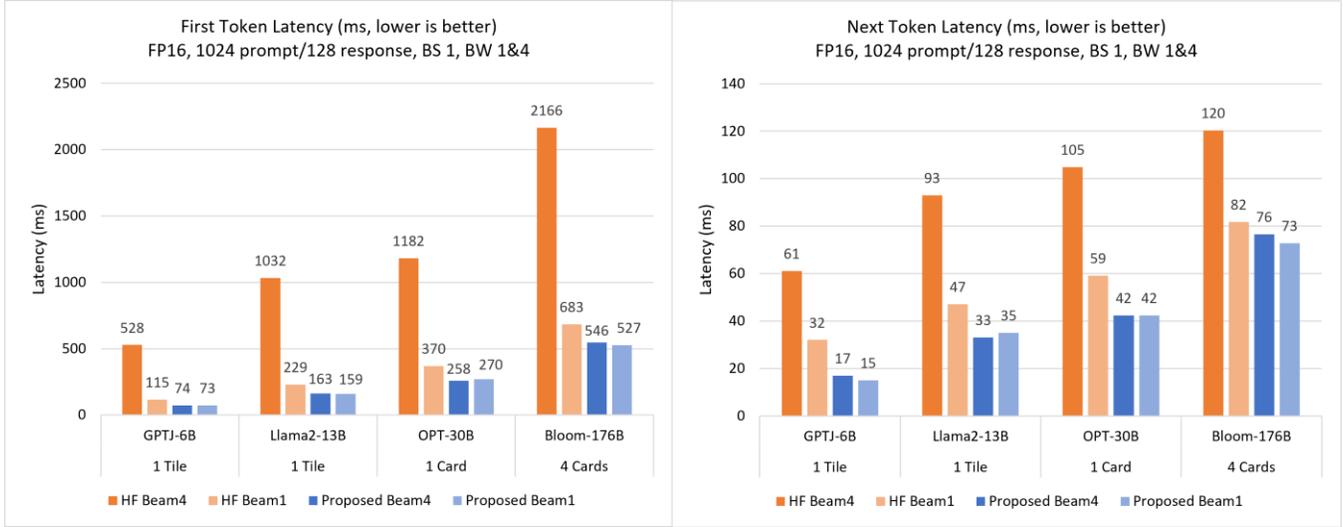

(a)                                                             (b)

Figure 8. (a) First and (b) next token latency performance of different LLMs with different parameters sizes

### *4.3 System Throughput Evaluation*

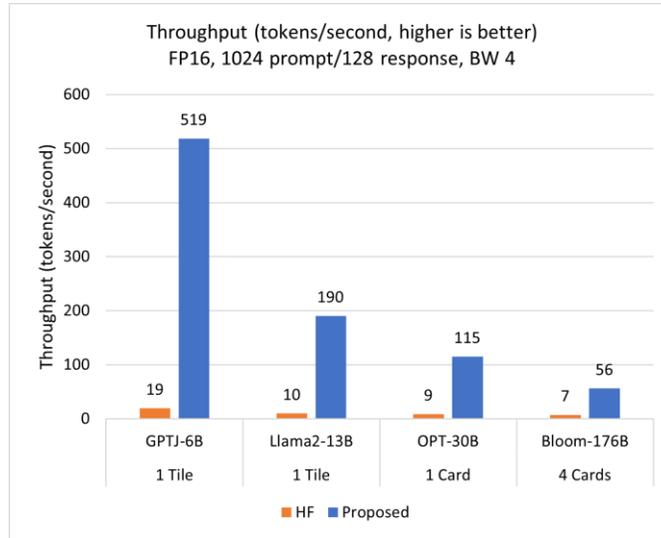

Figure 9. Throughput performance of different LLMs with different parameters sizes

We further compare our proposed LLM inference solution with standard HF implementation for throughput criteria computed by Eq (4). The performance results are shown in Figure 9, illustrating that the proposed method achieves 8x ~27x improvement on throughput. Besides the optimization for latency, the main benefit comes from our segment KV cache which can help save device memory then enlarge the runtime batch size value to improve the hardware resource utilization.

## 5. Conclusion

In this paper, we propose an efficient LLM inference solution with low latency and high throughput. Since LLM inference is a memory bound task, reducing the memory access frequency can help improve efficiency. The LLMs usually consist of multiple decoder layer, and we focus on simplifying the decoder layer structure by fusing data movement and element-wise operations to reduce memory access. Considering that LLM inference works at auto-regressive mode and consumes large device memory, we propose a segment KV cache policy to make the prompt being shared between different responses and regularly empty cache the fragments, so that to save device memory and improve the throughput. We also design an efficient SDPA kernel based on our fusion policy and the segment KV cache method. The proposed solution has been verified on several popular LLMs with parameter sizes from 6B~176B on Intel® GPU. The experimental results show that we achieve up to 7x lower latency and up to 27x higher throughput than the standard HuggingFace implementation.

## Acknowledgements

We much appreciate the contributions of Xiaodong Qiu, Cong li, Zhong Cao and other Intel® engineers to our work. We also want to give thanks to the technical guidance from Eric Lin, Fangwen Fu, Jiong Gong and other Intel® Architects. Thanks to the comments from Fan Zhao, Sophie Chen, Ting Ye, Kristina Kermanshahche and Brian Golembiewski to our paper.

## Notices and Disclaimers